\documentclass[prl,aps,twocolumn,groupedaddress,showpacs,floatfix,preprintnumbers]{revtex4}
\usepackage{amsmath,amssymb,multirow,epsfig,bm}

\newcommand{\etal}{{\it et al.,\;}}
\newcommand{\avg}[1]{\langle #1 \rangle}
\newcommand{\eqn}[1]{Eq.~(\ref{#1})}
\newcommand{\fig}[1]{Fig.~\ref{#1}}
\newcommand{\apriori}{\textit{a~priori} }              
\newcommand{\pictsize}{1.0}
\newcommand{\beq}{\begin{equation}}
\newcommand{\eeq}{\end{equation}}
\newcommand{\bea}{\begin{eqnarray}}
\newcommand{\eea}{\end{eqnarray}}
\newcommand{\pcut}{p_{\text{max}}}  
\newcommand{\omcut}{\omega_{\text{max}}} 
\newcommand{\eF}{\varepsilon_{F}}
\newcommand{\pF}{p_{F}}
\newcommand{\reff}{r_{\textrm{eff}}}

\begin{document}
\preprint{NT@UW-12-17}

\title{Cooper Pairing Above the Critical Temperature in a Unitary Fermi Gas}

\author{Gabriel Wlaz\l{}owski$^{1,2}$, Piotr Magierski$^{1,2}$, Joaqu\'{\i}n E. Drut$^3$, Aurel Bulgac$^{2}$ and Kenneth J. Roche$^{4,2}$}

\affiliation{$^1$Faculty of Physics, Warsaw University of Technology, Ulica Koszykowa 75, 00-662 Warsaw, Poland}
\affiliation{$^2$Department of Physics, University of Washington, Seattle, Washington 98195--1560, USA}
\affiliation{$^3$Department of Physics and Astronomy, University of North Carolina, Chapel Hill, NC 27599-3255 }
\affiliation{$^4$Pacific Northwest National Laboratory, Richland, WA 99352, USA}

\begin{abstract}
We present an {\it ab initio} determination of the spin response of the unitary Fermi gas. Based on finite temperature quantum Monte Carlo calculations and the Kubo linear-response formalism, we determine the temperature dependence of the spin susceptibility and the spin conductivity. We show that both quantities exhibit suppression above the critical temperature of the superfluid-to-normal phase transition due to Cooper pairing. The spin diffusion transport coefficient does not display a minimum in the vicinity of the critical temperature and drops to very low values $D_s\approx 0.8\,\hbar/m$ in the superfluid phase. All these spin observables show a smooth and monotonic behavior with temperature when crossing the critical temperature $T_c$, until the Fermi liquid regime is attained at the temperature $T^*$, above which the pseudogap regime disappears. 
\end{abstract}

\date{\today}

\pacs{03.75.Ss, 05.60.Gg, 51.20.+d, 05.30.Fk }

\maketitle

There are basically two flavors of superfluids - fermionic and bosonic.  The bosonic superfluid is realized when typically weakly interacting bosons condense and form a Bose-Einstein condensate (BEC). The typical fermionic superfluid is of Bardeen-Cooper-Schrieffer (BCS) type, where fermions in time-reversed orbitals form weakly bound Cooper pairs and turn into a BEC of Cooper pairs. The BEC superfluidity vanishes when the condensate fraction ceases to have a macroscopic value with increasing temperature and the long-range coherence is lost. On the other hand, when a BCS superfluid undergoes a phase transition to a normal state, Cooper pairs break due to thermal motion and there are no more bosonic constituents left to form a BEC. With the experimental confirmation of the BCS-BEC crossover in fermionic cold gases \cite{reviews,bcsbec}, it became clear that one can have a new kind of system where both bosonic and fermionic superfluid features are present at the same time. The paradigmatic example is the unitary Fermi gas (UFG) where, unlike the BCS or BEC cases, the interparticle interaction is strong, and the binding energy of the Cooper pair is comparable to the Fermi energy.  It is believed that the unpolarized unitary Fermi gas, in a temperature region just above the critical temperature $T_c$, exists in a state which is neither fully bosonic nor fully fermionic in character, called the pseudogap state, widely studied in high-$T_c$ superconductors (HTSC)~\cite{perali2,chen}. This is a temperature regime where a significant fraction of the Cooper pairs is present, even though the long-range coherence among them is lost, as is superfluidity. While the existence of the pseudogap regime in HTSC is an experimentally well known fact,  the nature of the corresponding regime around the critical temperature $T_c$ of the UFG has remained an open question. It is a tantalizing question, whether the pseudogap regime exists in dilute neutron matter as well. The properties of the neutron superfluid in the neutron star crust are very similar to those of the unitary gas, as highlighted by Bertsch's 1999 Many-Body X challenge. The reliable calculation of the neutrino processes in neutron stars, controlled by the neutron spin response \cite{neutrinos}, is a long-standing problem and the present results will likely shed new light on these phenomena. 

\begin{figure}[t]
\includegraphics[width=0.9\columnwidth]{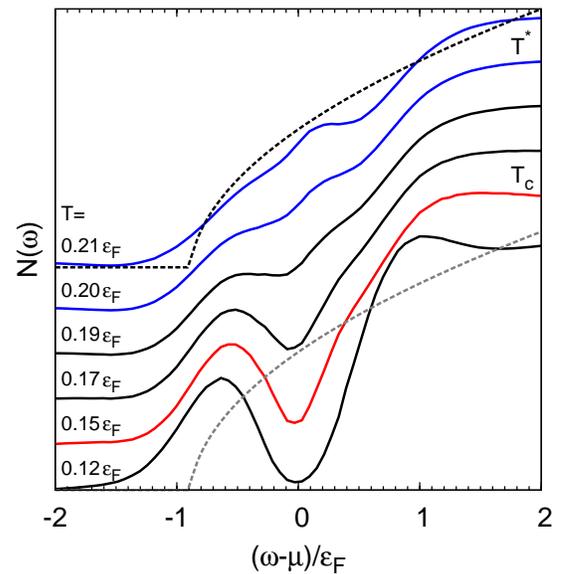}
\caption{ (Color online) Temperature evolution of the density of states profiles, extracted from the QMC simulations~\cite{mag2011}. The (blue) lines marked $T^{*}\approx0.2\,\eF$ correspond to the onset of the pseudogap regime,  the (red) line marked $T_c=0.15\,\eF$ corresponds to the critical temperature. 
The profiles for $T=0.21\,\eF$ and $T=0.12\,\eF$ are compared with the density of states for a Fermi liquid (dashed lines).
\label{fig:DOS} }
\end{figure}

The most striking feature of the pseudogap regime is the behavior of the fermionic density of states, which shows a dramatic depletion at the Fermi level. This was confirmed to exist in quantum Monte Carlo (QMC) simulations of the UFG \cite{mag2009,mag2011}, see Fig. 1, as well as in experiments \cite{stewart2010,gaebler2010,perali}. The transition from a pseudogap to a normal state in the UFG is somewhat similar to a gas-plasma transition, where no discontinuities are observed, which makes it difficult to observe. Indeed, no observable imprints on the thermodynamic properties have been detected in experiments so far \cite{salomon2,MIT_exp}, while at the same time theory also predicts none~\cite{mag2009,mag2011,BDM,dlwm}.  A direct measurement of the {\it local} density of states of a UFG in a trap is desirable. However, one can suggest different kinds of measurements as well. A strongly paired, but not necessarily superfluid, system would respond qualitatively differently to an external probe than a non-paired system, if one were to try to separate the two fermions in a pair. In particular, in a paired system the spin susceptibility and spin conductivity should be significantly suppressed when compared to the unpaired regime. One should therefore observe a marked difference between the response of a system in the pseudogap regime from a normal Fermi liquid one, which is the expected behavior at temperatures greater than $T^*$.  

Inspired by recent measurements of various spin responses like the spin susceptibility~\cite{Sanner}, spin transport coefficients~\cite{Sommer} or dynamic (spin) structure factors~\cite{Hoinka}, we present here an {\it ab initio} evaluation of the spin susceptibility $\chi_s$ and the spin conductivity $\sigma_s$ for unpolarized homogeneous unitary Fermi gas. We show that both of these quantities carry a strong signal indicating survival of the Cooper pairs above the critical temperature.  Our new results are consistent with previous studies~\cite{mag2009,mag2011}, where the existence of the pseudogap regime was found between the critical temperature $T_c=0.15(1)\,\eF$ and $T^{*}=0.19(2)\,\eF$, where $\eF=\pF^2/2m$ is the free Fermi gas energy, and $\pF=\hbar\,(3\pi^{2}n)^{1/3}$ is the Fermi momentum corresponding to the total particle number density $n$. While in the previous estimation of the $T^*$ temperature finite resolution  of the method provided only the lower bound, the analysis of the spin responses allows us to provide a significantly more accurate interval for the pseudogap onset.

Moreover, the computed responses allow us to extract the temperature dependence of the spin diffusion coefficient from first principle calculations.  There has been considerable speculation as to whether the transport coefficients possess universal lower bounds imposed by quantum mechanics. The best known example is a conjecture formulated by  Kovtun, Son, and Starinets of the existence of a lower bound $\eta/s\geqslant\hbar/(4\pi k^{}_{B})$  on the ratio of the shear viscosity $\eta$ to the entropy density $s$ for all fluids~\cite{KSS}. While in our previous work~\cite{Wlazlowski} we found that path integral Monte Carlo (PIMC) calculation is compatible with a well-defined minimum for the $\eta/s$ ratio in the vicinity of the critical temperature, here we show that the spin diffusion does not exhibit a similar minimum as a function of temperature. 

To determine the spin properties of the UFG we employ the PIMC technique on the lattice, which provides numerically exact results, up to quantifiable systematic uncertainties (for details see Ref.~\cite{BDM}). Since the pseudogap regime is expected to exist in a rather small temperature region $0.15\lesssim T/\eF\lesssim0.2$ it needs to be checked if it survives when the thermodynamic $V \to \infty$ and continuum $n \to 0$ limits are recovered. This step is of great importance as the critical temperature approaches the value $T_c=0.15(1)\,\eF$ only in the thermodynamic and continuum limits~\cite{BDM}. At the  same time the temperature $T^*$ above which there are no Cooper pairs left, and which reflects short range correlations among particles, does not show a similar strong volume dependence. To check the stability of the results, as the thermodynamic and continuum limit are approached,  we performed simulations using three lattices $N_x=8,10,12$ with corresponding average density $n\backsimeq0.08, 0.04$, and $0.03$, respectively. The systematic errors related to finite volume effects as well as effective-range corrections, are estimated to be likely $\sim 10\%-15\%$, while the statistical errors of the PIMC data are below $1\%$; see~\cite{Supplemental} for more details.  Henceforth, we define units $\hbar=m=k_{B}=1$.

The spin susceptibility as well as the spin conductivity can be theoretically determined using linear response theory via the Kubo relations. The uniform static spin susceptibility  $\chi^{}_s=\partial(n^{}_{\uparrow}-n^{}_{\downarrow})/\partial(\mu^{}_{\uparrow}-\mu^{}_{\downarrow})$  is obtained as~\cite{Mahan} 
\begin{equation}
\chi^{}_s = \lim_{q \to 0}\dfrac{1}{V}\int_{0}^{\beta}d\tau
\avg{\hat{s}_{\bm{q}}^{z}(\tau)\hat{s}_{-\bm{q}}^{z}(0)},
\label{eqn:chi_s}
\end{equation} 
where $\hat{s}_{\bm{q}}^{z}=\hat{n}_{\bm{q}\uparrow}-\hat{n}_{\bm{q}\downarrow}$ represents  a difference  between spin-selective particle number operators in Fourier representation  $\hat{n}_{\bm{q}\lambda}= \sum_{\bm{p}}\hat a^\dagger_\lambda(\bm{p}) \, \hat a_\lambda^{}(\bm{p}+\bm{q})$, $\beta=1/T$ is the inverse  temperature and $\avg{\ldots}$ stands for the grand-canonical ensemble average. The imaginary-time dependence of an operator is generated as $\hat{O}(\tau)=e^{\tau(\hat{H}-\mu\hat{N})}\hat{O}e^{-\tau(\hat{H}-\mu\hat{N})}$, where $\hat{H}$ is the Hamiltonian of the system, $\mu$ is the chemical potential, and $\hat{N}$ is the particle number operator.  The expectation values can be evaluated directly for $q=0$. In this case the spin operator $\hat{s}_{\bm{q}=0}^{z}$ commutes with the Hamiltonian and the expectation value is $\tau$ independent. Consequently, the QMC calculation consists of the evaluation of the expectation value of a two-body operator, and the static spin susceptibility can be computed very accurately within our framework.

\begin{figure}
\includegraphics[width=\pictsize\columnwidth]{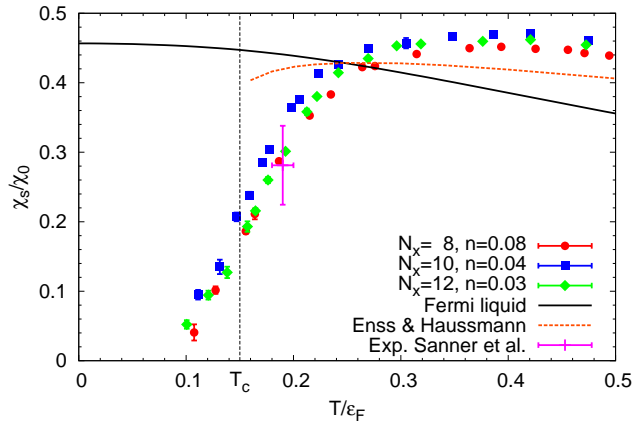}
\caption{ (Color online) The static spin susceptibility as a function of temperature for an $8^3$ lattice solid (red) circles, $10^3$ lattice (blue) squares and $12^3$ lattice (green) diamonds. The vertical black dotted line indicates the critical temperature of the superfluid to normal phase transition $T_c=0.15\,\eF$. For comparison, the Fermi liquid theory prediction and recent results of the $T$-matrix theory produced by Enss and Haussmann~\cite{Enss} are plotted with solid and dashed (brown) lines, respectively. The experimental data point from Ref.~\cite{Sanner} is also shown. \label{fig:1} }
\end{figure}
In \fig{fig:1}, the static spin susceptibility $\chi^{}_s$ in units of free Fermi gas susceptibility $\chi^{}_0=3n/2\eF$ is shown for temperature range $0.1\leqslant T/\eF\leqslant0.5$. The results on $8^{3}$, $10^{3}$, and $12^{3}$ lattices exhibit satisfactory agreement with each other and no systematic trend in the data has been detected as we approach the thermodynamic and the continuum limit. For temperatures $0.25\,\eF-0.5\,\eF$ no strong temperature dependence of the spin susceptibility is observed. The latter is well below the susceptibility of the free Fermi gas with a value around $\chi^{}_s\approx0.45\,\chi^{}_{0}$, which is in qualitative agreement with the Fermi liquid picture as well as results of other groups~\cite{Wulin,Palestini,Kashimura,Enss,Mink}. In the interval $T^*=0.20-0.25\,\eF$ we find the beginning of pronounced suppression of the spin susceptibility, which we associate with the existence of Cooper pairs in the system. Note that already at $T_c$ the spin susceptibility is about half its value at the onset of suppression (roughly $T=0.25\,\eF$). Thus, two temperature scales are clearly distinguishable: the critical temperature of the superfluid-to-normal phase transition $T^{}_c=0.15\,\eF$, and the onset of the Cooper-pair formation $T^*$.

\begin{figure}
\includegraphics[width=\pictsize\columnwidth]{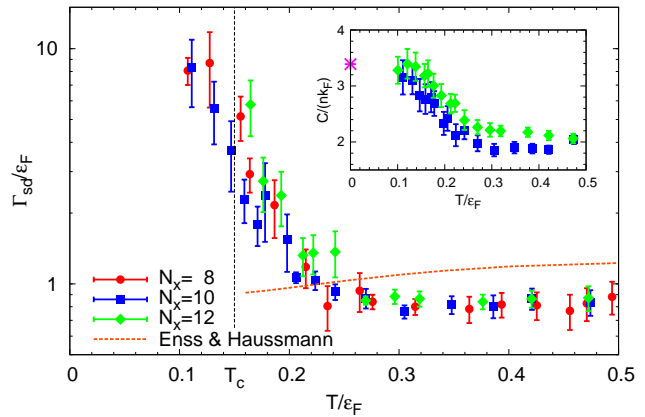}
\caption{ (Color online)  The spin drag rate $\Gamma_{sd}=n/\sigma_s$ in units of Fermi energy as a function of temperature for an $8^3$ lattice solid (red) circles, $10^3$ lattice (blue) squares and $12^3$ lattice (green) diamonds. Vertical black dotted line locates the critical temperature of superfluid to normal phase transition. Results of the $T$-matrix theory are plotted by dashed (brown) line~\cite{Enss}. The inset shows the extracted value of the contact density as function of the temperature. The (purple) asterisk shows the contact density from the QMC calculations of Ref.~\cite{Hoinka2} at $T = 0$.
\label{fig:2} }
\end{figure}
The static spin conductivity $\sigma_{s}$ represents another quantity which is expected to be strongly affected by the presence of the Cooper pairs as it measures response of the spin current $\bm{j}_{s}=\bm{j}_{\uparrow}-\bm{j}_{\downarrow}$ once the weak external $\bm{F}$ force which couples with opposite signs to the two spin populations is applied to the system, i.e., $\bm{j}_{s}=\sigma_s\bm{F}$. In order to extract the spin conductivity we consider the Kubo formula, which relates the frequency-dependent spin conductivity to the corresponding spectral density: $\sigma_{s}(\omega)=\pi\rho^{(jj)}_{s}(\bm{q}=0,\omega)/\omega$; while the static spin conductivity is defined in the limit of zero frequency: $\sigma_{s}=\lim_{\omega\rightarrow 0^{+}}\sigma_{s}(\omega)$.  The spectral density $\rho^{(jj)}_{s}(\bm{q},\omega)$ is related to the imaginary-time (Euclidean) current-current correlator provided by the PIMC method
 \begin{equation}
  G^{(jj)}_{s}(\bm{q},\tau)=\dfrac{1}{V}  \avg{
  [ \hat{j}^{z}_{\bm{q}\uparrow}(\tau)-\hat{j}^{z}_{\bm{q}\downarrow}(\tau)]
  [ \hat{j}^{z}_{\bm{-q}\uparrow}(0)-\hat{j}^{z}_{\bm{-q}\downarrow}(0)]
  },
 \end{equation}
by inversion of the spectral relation
\begin{equation}
G^{(jj)}_{s}(\bm{q},\tau)=\int_{0}^{\infty}\rho^{(jj)}_{s}(\bm{q},\omega)\,\dfrac{\cosh\left[ \omega(\tau-\beta/2)\right] }{\sinh\left( \omega\beta/2\right) }\,d\omega,
 \label{eqn:AnalitycalContinuation}
 \end{equation} 
where $\hat{j}^{z}_{\bm{q}\lambda}(\tau)$ stands for the third component of Fourier representation of the spin-selective current operator. To invert \eqn{eqn:AnalitycalContinuation} we have applied the methodology which combines two complementary methods: singular value decomposition and maximum entropy method, both described in Ref.~\cite{MagierskiWlazlowski}.  An additional \apriori information includes the non-negativity of the spin conductivity $\sigma_{s}(\omega)\geqslant 0$, a Lorentzian-like structure at low frequencies (Drude model), and the asymptotic tail behavior $\sigma_{s}(\omega\to\infty)=C/(3\pi\omega^{3/2})$, where $C$ is the Tan contact density~\cite{Enss}. The contact density was extracted from the tail of the momentum distribution $n(p)\sim Cp^{-4}$, using a technique similar to that of Ref.~\cite{Drutetal2}.  As the universal decay of the tail distribution starts around $p/\pF\approx 2$,  we were unable to extract the contact for the $N_x=8$ lattice. In that case we used the value of $C$ extracted from the $N_x=10$ lattice.  The inset of \fig{fig:2} shows the temperature dependence of the contact density used as \apriori information. For calculations in the $N_x=12$ lattice, we found that the signal-to-noise ratio for the correlators at $T<0.16\,\eF$ is insufficient to perform a stable reconstruction of the spectral density.  For more details about the reconstruction process see the Supplemental Material~\cite{Supplemental}.

In \fig{fig:2} we present the inverse of the static spin conductivity called the spin drag rate $\Gamma_{sd}=n/\sigma_s$, which is the rate of the momentum transfer between fermions with opposite spins.\ For $T = 0.25\,\eF-0.5\,\eF$, no strong temperature dependence is observed. For all three lattices the spin drag rate exhibits a significant enhancement above $T_c$, in the interval $T^*=0.20-0.25\,\eF$, which is consistent with the occurrence of the spin susceptibility suppression. Such an enhancement is expected for a system with strong correlations between particles of opposite spins. 

\begin{figure}
\includegraphics[width=\pictsize\columnwidth]{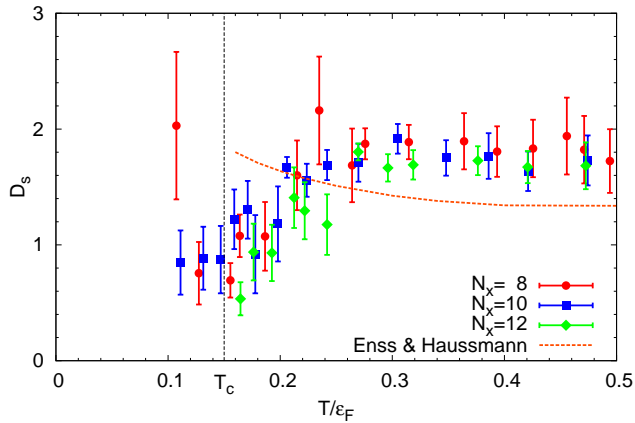}
\caption{ (Color online) The spin diffusion coefficient obtained by the Einstein relation $D_s=\sigma_s/\chi_s$ as function of temperature. The notation
is identical to \fig{fig:2}.
\label{fig:3} }
\end{figure}
Finally, the computed spin susceptibility and spin conductivity allow us to extract the spin diffusion coefficient $D_s$ in a fully {\it ab initio} manner. In the hydrodynamic regime it defines the proportionality between the spin current $\bm{j}_s$ and spatially varying polarization by Fick's law $\bm{j}_s=-D_{s}\bm{\nabla}(n_{\uparrow}-n_{\downarrow})$. The spin diffusivity  $D_s$ can be related to the spin conductivity and the spin susceptibility by the Einstein relation $D_s=\sigma_s/\chi_s$.  In \fig{fig:3} we show the temperature evolution of the spin diffusion coefficient. In the normal phase, for temperatures $0.25\,\eF-0.5\,\eF$ the diffusivity is approximately constant $D_s\approx 1.8$. Surprisingly, we find that the spin diffusion coefficient decreases substantially when the system enters the pseudogap regime, acquiring eventually a value around $D_s\approx 0.8$ in the superfluid phase. Such a low value can be understood as a quantum limit for this transport coefficient. The bound originates from kinetic theory, where $D_s\sim vl$, $v$ is the average particle speed, and $l$ is the mean free path. For a strongly correlated system, the product of $v\sim \pF\sim n^{1/3}$ and $l\sim n^{-1/3}$ cancels the density dependence, giving $D_s\sim 1$.

In Ref.~\cite{Bruun} the existence of a minimal value of the diffusivity was predicted for a temperature somewhat below the Fermi temperature, within Landau-Boltzmann theory. Recently, it was reported that Luttinger-Ward theory sets the minimum $D_s\backsimeq 1.3$ at $T=0.5\,\eF$~\cite{Enss}. Our {\it ab initio} calculations do not confirm the presence of a minimum for the spin diffusion coefficient down to $T=0.1\,\eF$, and they do not rule out possibility that the diffusivity $D_s$ decreases further when temperature is lowered.

Our results for the spin susceptibility and the spin drag rate deviate from the recent measurements of the MIT group~\cite{Sommer} extracted from fully polarized cloud collisions.  This technique, in which two noninteracting clouds collide, and in which pairs do not exist and likely are not formed, is less suitable to probe the low temperature regime, where pairs already exist and their presence is of crucial importance. Pair formation in such an experiment would require three-body collisions. Recent theoretical simulations  \cite{Palestini,Taylor,Goulko} demonstrated that this experiment can be explained assuming that the measurement explores a nonequilibrium state associated with a quasirepulsive Fermi gas and two-body collisions alone. On the other hand, the technique based on speckle imaging of spin fluctuations, used to measure the spin susceptibility of the system in thermal equilibrium~\cite{Sanner}, is in remarkable agreement with our theoretical results; see \fig{fig:1}.

In summary, we have presented results for the spin response of the UFG at finite temperature, obtained through an
{\it ab initio} PIMC approach. The spin susceptibility and the spin conductivity bear signatures of Copper-pair formation above the critical temperature $T_c\backsimeq0.15\,\eF$ up to $T^{*}\approx0.20-0.25\,\eF$. The spin diffusion coefficient does not display a minimum in the vicinity of the critical temperature, but instead drops to very low values $D_s\approx 0.8$ in  the superfluid phase. We showed that the spin response of a unitary Fermi gas is not affected by the superfluid to normal transition, but only by the presence of Cooper pairs, and all these spin observables show a smooth and monotonic behavior up to the temperature $T^*$, where the pseudogap cease to show up in the density of states.

We thank T. Enss for providing us with the $T$-matrix results~\cite{Enss}. We acknowledge support under U.S. DOE Grant No. DE-FG02-97ER41014 and No. DE-FC02-07ER41457, and Contract No. N N202 128439 of the Polish Ministry of Science. One of the authors (G.W.) acknowledges the Polish Ministry of Science for the support within the program ``Mobility Plus \!-\! I edition'' under Contract No. 628/MOB/2011/0. 
Calculations reported here
have been performed at the Interdisciplinary Centre for
Mathematical and Computational Modelling (ICM) at
Warsaw University and on the University of Washington
Hyak cluster funded by the NSF MRI Grant No. PHY-0922770.
This research also used resources of the National Center 
for Computational Sciences at Oak Ridge National Laboratory, 
which is supported by the Office of Science of the Department 
of Energy under Contract DE-AC05-00OR22725.

{\it Note added.}  Similar studies of $\chi_s(T)$ and density of states in an attractive 2D Hubbard model (in the context of cuprates) have been performed in Refs. \cite{mohit}. There are, however, qualitative differences between the physics of the attractive 2D Hubbard model and dilute Fermi gases in two dimensions, which are practically noninteracting; see Chap. 7 in Ref. \cite{bcsbec}.
 

\begin{center}
{\bf Supplemental online material for:}\\
{\bf ``Cooper Pairing Above the Critical Temperature in a Unitary Fermi Gas''}\\
\end{center}

\begin{small}
\noindent
This supplemental material provides discussion of the systematic errors
and technical details concerning the inversion procedure.
\end{small}
\setcounter{equation}{0}
\setcounter{figure}{0}

\section{Systematic errors}

\begin{figure}
\includegraphics[width=\pictsize\columnwidth]{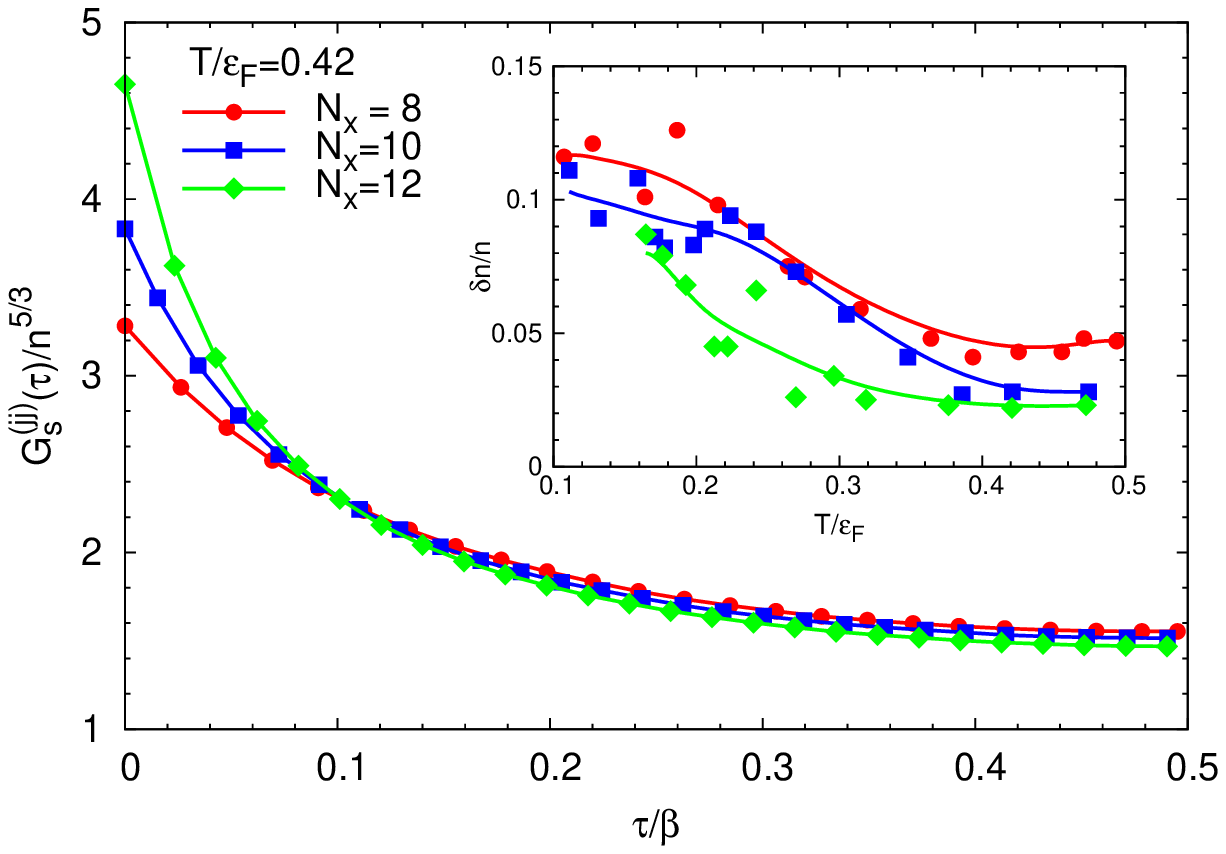}
\caption{ (Color online) The dimensionless current-current correlator for temperature $T=0.42\,\eF$ as function of the imaginary time for $8^3$ lattice solid (red) circles, $10^3$ lattice (blue) squares and $12^3$ lattice (green) diamonds. The inset shows violation of the spin conductivity sum rule in the lattice simulations as the function of temperature. Solid lines are a guide to the eye.
\label{fig_sup:1} }
\end{figure}

The results of lattice simulations are affected by two types of errors: statistical and systematic. The former are explicitly under control as we collected ensemble of $10^4$ uncorrelated samples at each temperature, reducing the statistical uncertainties below $1\%$. The systematic errors arise due to the finite volume of the box $V=N_{x}^{3}$ (we set lattice spacing to unity)  and the finite density $n$ of the system. The latter is related to corrections coming from the nonzero effective range $\reff$ and exclusion of the universal high momenta tail in the occupation probability $n(p)\sim Cp^{-4}$. In the continuum limit $n\to 0$  both corrections are negligible as $\pF\reff\to 0 $ and $\pcut/\pF\to +\infty$, where $\pcut=\pi$ is the maximum momentum  available on the lattice. In our simulations we have reached values $\pF\reff\backsimeq 0.54,\,0.43,\,0.39$ and $\pcut/\pF\backsimeq 2.4,\,3.0,\, 3.3$, respectively for lattices $N_x=8,\,10$ and $12$. In particular  the moderate value of $\pcut/\pF$ leads to significant bias of the current-current correlator $G^{(jj)}_{s}$. It is easy to note that contribution from the universal spin conductivity tail $\sigma_{s}(\omega\to\infty)=C/(3\pi\omega^{3/2})$  leads to $G^{(jj)}_{s}(\tau=0)=+\infty$. This condition is violated in PIMC simulations. In \fig{fig_sup:1} we present the evolution of the $G^{(jj)}_{s}(\tau)$ correlator for temperature $T=0.42\,\eF$ when the continuum limit is approached. The influence of the correlator by the systematic errors for low imaginary times is clearly visible.  Note however, that the most valuable information about the low frequency part of the spin conductivity  comes from the values of the correlator located around $\tau/\beta=1/2$, where the contribution from the universal tail is marginal.

In Refs~\cite{EnssSupp,Enss2Supp} it was shown that the spin $f$-sum rule implies ($m=1$)
\begin{equation}
 \int_{-\infty}^{+\infty}\dfrac{d\omega}{\pi}\sigma_{s}(\omega)=n.
 \label{eqn:spinfSumRule}
\end{equation} 
It is straightforward to prove that
\begin{equation}
  \int_{-\infty}^{+\infty}\dfrac{d\omega}{\pi}\sigma_{s}(\omega)=\int_{0}^{\beta}G^{(jj)}_{s}(\tau)\,d\tau,
\end{equation} 
and thus the sum rule~(\ref{eqn:spinfSumRule}) can be checked directly in the lattice simulations and used  as an approximate measure of the systematic errors. The deviation $\delta n=n-\int_{0}^{\beta}G^{(jj)}_{s}(\tau)\,d\tau$ from the expected result is presented in the inset of the \fig{fig_sup:1}. We have found that the discrepancy does not exceed $15\%$ for each case, and the strongest violation of the sum rule is in the low temperature regime. This is in agreement with the previous estimations of the systematic errors reported in Ref.~\cite{BDMSupp}. Moreover, we observe that the sum rule is typically underestimated ($\delta n>0$). This fact as well as the finite value of the correlator $G^{(jj)}_{s}(\tau=0)$ are due to the existence of the cut-off in the momentum space which implies the existence of the cut-off in the frequency space at $\omcut/\eF\approx (\pcut/\pF)^2$.

Our simulations predict the temperature of Cooper-pair formation to be $T^*/\eF\approx0.20-0.25$, which does not change significantly with the box size and the density (no systematic trend has been detected). On the other hand, the critical temperature $T_c$ does show finite-range effects. One may therefore ask the following question: Can the systematic errors merge the two distinct temperatures $T_c$ and $T^*$ into one temperature?  Since the temperature of 
pair formation is not visibly influenced by the systematic corrections, we assume that $T^*$ is not lower than $0.2\,\eF$ in the continuum limit. In order to shift the critical temperature $T_c=0.15\,\eF$ to $T^*\approx0.2\,\eF$, one should assume that the systematic errors generated by effective-range corrections are at least $30\%$, which is far above our estimations. Moreover, results reported in Refs.~\cite{BurovskiSupp,Burovski2Supp,Chen2Supp} (see also Ref.~\cite{Wingate1}) do not reveal such 
strong departure of $T_c$ from its continuum limit value for densities considered in this work. A more precise analysis, utilizing improved actions and operators such as those constructed in Refs.~\cite{Kaplan,JED}
is required to rigorously quantify the uncertainties, but this is beyond the scope of this work.

Recently, Castin and Werner \cite{CastinWernerSupp} made the claim that on a lattice, the manner the ultraviolet cutoff is implemented might lead to corrections which do not vanish when the continuum limit is taken, especially if one adopts a so-called spherical momentum cutoff. 
 
In lattice calculations one introduces a single-particle propagator with an ultraviolet momentum cutoff
 \begin{equation}
 \label{eq:G1}
 G_1(\bm{r},t)=\int_{-\infty}^{\infty} \frac{d\omega}{2\pi i}\int_D\frac{d^3\bm{k}}{(2\pi)^3} 
 \frac{ \exp(i\bm{k}\cdot\bm{r}-i\omega t) }{ \omega+i0^+-\varepsilon(\bm{k})},
 \end{equation} 
where   $\varepsilon(\bm{k})=\hbar^2k^2/2m$, $\bm{r}=(n_x,n_y,n_z)l$ stand for discrete lattice 
points and  where $l$ is the lattice constant, the domain of integration $D$ is either $|\bm{k}|\leq \Lambda =\pi/l$ (spherical momentum cutoff) or $|k_x|\le \Lambda$,  $|k_y|\le \Lambda$, $|k_z|\leq \Lambda$ (cubic cutoff). If the lattice is infinite $n_{x,y,z}$ run over all integer numbers. In the case of a finite box with periodic boundary conditions $n_{x,y,z}$ run over a finite interval only and in this case the integral becomes a discrete sum in which $d^3\bm{k}$ is replaced by $(2\pi)^3/V$, where $V$ is the volume of the box. From the single-particle propagator one then constructs the two-particle propagator $G_1(\bm{r}_1-\bm{r}_2,t)G_1(\bm{r}_1-\bm{r}_2,t)$ assuming that the two particles have to be at the same lattice point in order to interact.  In the energy momentum representation this two-particle propagator
reads
\begin{eqnarray}
&& G_2(\bm{K},E)= \nonumber \\
&& \!\!\!\!\! \int_{D_{12}} \frac{d^3\bm{q}}{(2\pi)^3}
\frac{1}{E+i0^+ -\varepsilon(\bm{K}/2+\bm{q})-\varepsilon(\bm{K}/2-\bm{q})}, \label{eq:G2}
\end{eqnarray} 
where $D_{12}=D_1\cap D_2$ is the intersection of the two-momentum domains defined above either for a cubic or spherical cutoff for each single-particle momentum $\bm{K}/2\pm \bm{q}$ respectively, and $\bm{K}$ is center of mass momentum of the pair.
Usually the strength of the on-site interaction between two particles is obtained by assuming that both $E$ and $\bm{K}$ vanish in the Lippmann-Schwinger equation, namely (we use convention $V(\bm{r}_1-\bm{r}_2)=-g_0\delta(\bm{r}_1-\bm{r}_2)$)
\begin{equation}
-\frac{1}{g_0} = \frac{m}{4\pi\hbar^2a} - \int_D \frac{d^3\bm{q}}{(2\pi)^3}\frac{1}{2\varepsilon(\bm{q})}=\frac{m}{4\pi\hbar^2a} - \frac{m\Lambda}{2\pi^2\hbar^2}.
\end{equation}
Notice than in this case $D_{12}\equiv D$. In the case of a cubic momentum cutoff the coefficient in front of the last term is slightly different \cite{CastinWernerSupp}. If the interacting pair has a finite center of mass momentum $\bm{K}$ and total energy $E=\hbar^2K^2/4m+\hbar^2k^2/m$, the corresponding Lippmann-Schwinger equation reads 
\begin{eqnarray}
&&-\frac{1}{g} = \frac{1}{t(\bm{K},E)} + \frac{m}{\hbar^2}\int_{D_{12}} \frac{d^3\bm{q}}{(2\pi)^3} \frac{1}{k^2+i0^+ -q^2}\nonumber \\
&& =\frac{m}{4\pi\hbar^2} \left ( \frac{1}{a} +ik \right ) -\frac{m\Lambda}{2\pi^2\hbar^2}- \frac{ikm}{4\pi\hbar^2} - \nonumber \\
&& \frac{ m}{2\pi^2\hbar^2} \int_0^\Lambda q_\perp dq_\perp 
 \int_{\sqrt{\Lambda^2-q_\perp^2}-|K|/2} ^{\sqrt{\Lambda^2-q_\perp^2} } \frac{dq_x}{k^2 +i0^+ -q_\perp^2-q_x^2} \nonumber \\
&&\approx\frac{m}{4\pi\hbar^2a}  -\frac{m\Lambda}{2\pi^2\hbar^2} +\frac{m|K|}{8\pi^2\hbar^2}
\end{eqnarray}
where $t(\bm{K},E)$ is $t$-matrix of the pair.
We neglected terms of order $k^2$ and $K^2$ in evaluating the above integrals. Now we introduce an on-site coupling constant $g$ for each value of the center of mass energy/momentum
\begin{eqnarray}
-\frac{1}{g} = \frac{m}{4\pi\hbar^2a}  - \frac{m\Lambda}{2\pi^2\hbar^2} +\frac{m|K|}{8\pi^2\hbar^2}  \nonumber \\
=\frac{m}{4\pi\hbar^2a}- \frac{m\Lambda}{2\pi^2\hbar^2}\left (1- \frac{|K|}{4\Lambda} \right ).
\end{eqnarray}
In a unitary Fermi gas the average value of $|K|$ is of the order of $k_F$, and thus this amounts to a correction of the order of $k_F/\Lambda$, which vanishes in the continuum limit. Notice that by defining in this manner $g$ the pole of the $t$-matrix is always at $k=i/a$ irrespective of the center of mass momentum of the pair. At unitarity one has
\begin{equation}
g= g_0 \left (1-\frac{|K|}{4\Lambda}\right )^{-1} \approx g_0  + \frac{g_0}{4} {O} \left(\frac{k_F}{\Lambda} \right ),
\end{equation}
thus a relatively small and systematic correction to the coupling constant, which can be safely evaluated in perturbation theory and which vanishes in the continuum limit, when $\Lambda \rightarrow \infty$. In our simulations we neglect this correction to the coupling constant and the results correspond to an interaction strengths shifted slightly towards the BCS limit.

Another lattice artifact is the undesired finite range of the order of the lattice constant, which the contact interaction acquires. There have been in literature a number of attempts to remedy this deficiency, one of the most popular being the modification of the dispersion relation of the kinetic energy in various ways, see Ref. \cite{CastinWernerSupp} and a number of references therein to earlier works. These suggestions amount to modifying the single-particle kinetic energy used in the two-particle propagator (\ref{eq:G2}) such that
\begin{equation}
\varepsilon(\bm{k}) = \frac{\hbar^2 \bm{k}^2}{2m} \to  \frac{\hbar^2 \bm{k}^2}{2m}
\left ( 1+\alpha \frac{\bm{k}^2}{\Lambda^2}+\ldots \right), 
\end{equation}
where the parameters $\alpha$, $\Lambda$ and possible higher order terms are chosen so as to either minimize the effective range, or even make it exactly zero, when such a dispersion relation is used for the two-particle propagator (\ref{eq:G2}) in the Lippmann-Schwinger equation. In particular, for $\alpha >0$ the wave function of the two-particle bound state would have the form $[\exp(-br)-\exp(-cr)]/r$ (with $b>0$ and $c>0$) instead of the expected $\exp(-r/a)/r$, and the $1/r$ singularity at the origin disappears. The negative effect of such a procedure is that the expectation value of the kinetic energy of the gas is modified as well and the properties of the free Fermi gas are not reproduced correctly in the corresponding QMC calculation.  Since the single-particle momentum distribution at large momenta does not satisfy anymore the expected behavior $n(k) \propto C/k^4$ all Tan relations \cite{TanSupp} are violated in the entire BCS-BEC crossover interval at all temperatures. The energy of the interacting system would not satisfy anymore the expected high-temperature behavior $E(T) \rightarrow 3TN/2 +\ldots $ for $T\rightarrow \infty$ and the virial expansion would be violated as well.

\section{Analytic continuation \& inversion}
To compute the frequency dependent spin conductivity one has to invert the integral equation, for zero momentum transfer $\bm{q}=0$:
\begin{equation}
 G^{(jj)}_{s}(\tau)=\int_{0}^{\infty}\rho^{(jj)}_{s}(\omega)\,K(\omega,\tau)\,d\omega,
 \label{eqn:AnalitycalContinuationSupp}
 \end{equation} 
 with the kernel
\begin{equation}
 K(\omega,\tau)=\dfrac{\cosh\left[ \omega(\tau-\beta/2)\right] }{\sinh\left( \omega\beta/2\right) }.
\end{equation} 
Analogously to the extraction of the shear viscosity (see in our previous work~\cite{WlazlowskiSupp}),  it is convenient to rewrite the \eqn{eqn:AnalitycalContinuationSupp} in the form:
\begin{equation}
  G^{(jj)}_{s}(\tau)=\dfrac{1}{\pi}\int_{0}^{\omcut}\sigma_{s}(\omega)\,\tilde{K}(\omega,\tau)\,d\omega,
  \label{eqn:InversionProblemSupp}
\end{equation} 
where $\tilde{K}(\omega,\tau)=\omega\,K(\omega,\tau)$ and $\sigma_{s}(\omega)=\pi\rho^{(jj)}_{s}(\omega)/\omega$. Note that we have introduced explicitly the frequency cut-off (see discussion in the previous section).
The correlator $G^{(jj)}_{s}(\tau)$ is provided by the PIMC method for at least 51 points in imaginary time $\tau$, uniformly distributed in the interval $[0,\beta]$. As the correlator satisfy $G^{(jj)}_{s}(\tau)=G^{(jj)}_{s}(\beta-\tau)$ we restrict the inversion to the interval $\tau\in[0,\beta/2]$, and apply the symetrization: $G^{(jj)}_{s}(\tau)\leftarrow[G^{(jj)}_{s}(\tau)+G^{(jj)}_{s}(\beta-\tau)]/2$ in order to decrease statistical fluctuations.

\begin{figure}
\includegraphics[width=\pictsize\columnwidth]{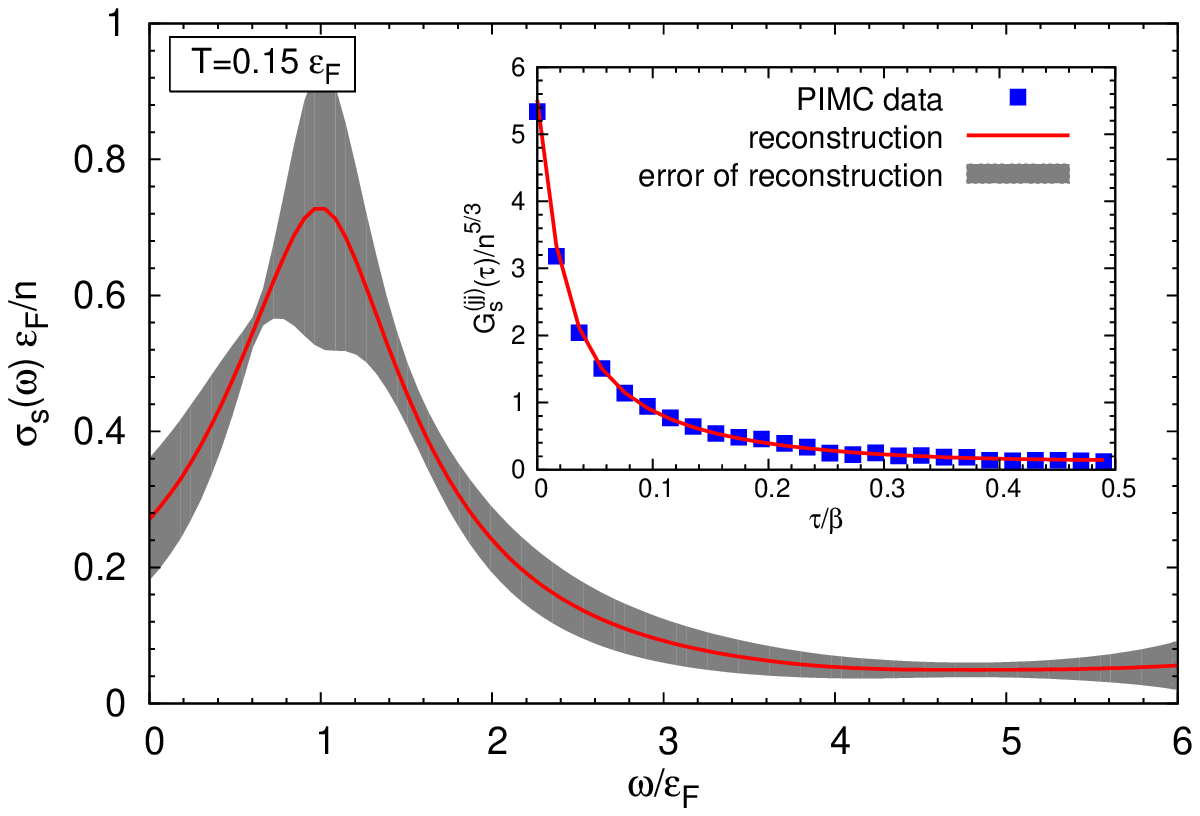}
\includegraphics[width=\pictsize\columnwidth]{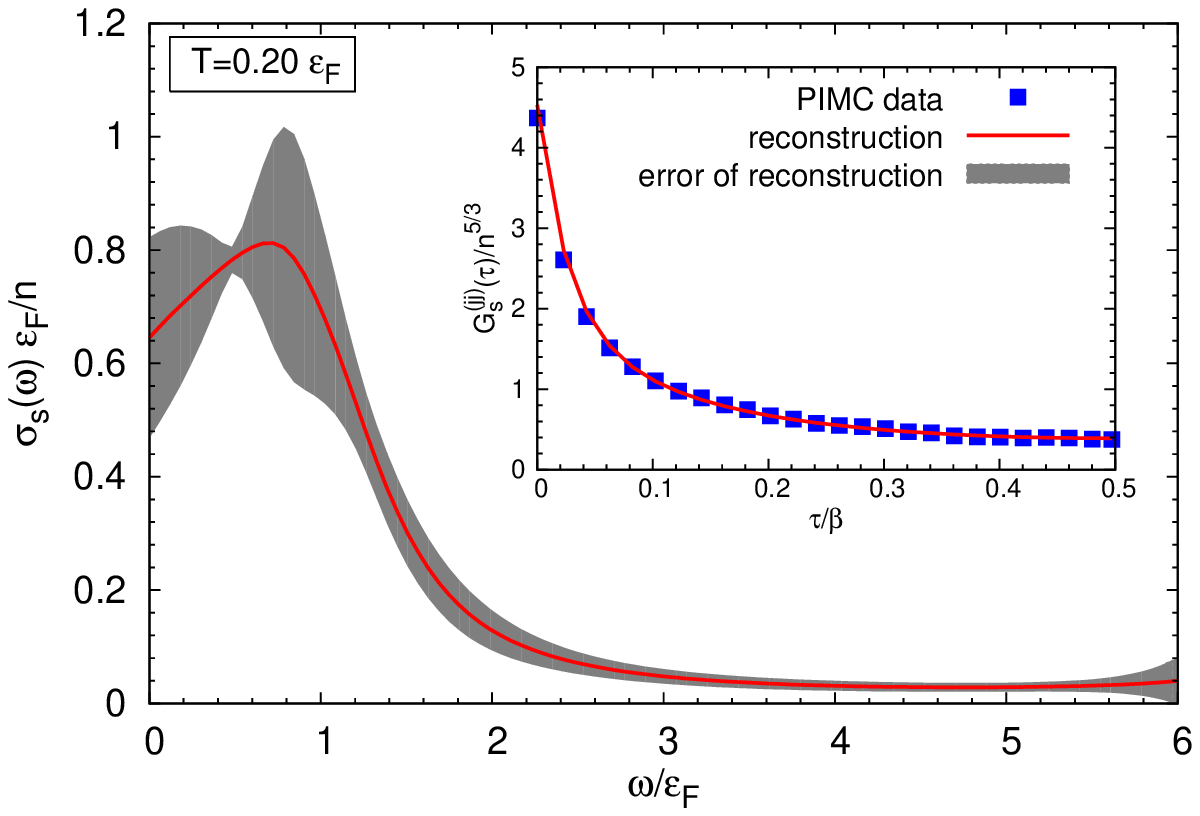}
\includegraphics[width=\pictsize\columnwidth]{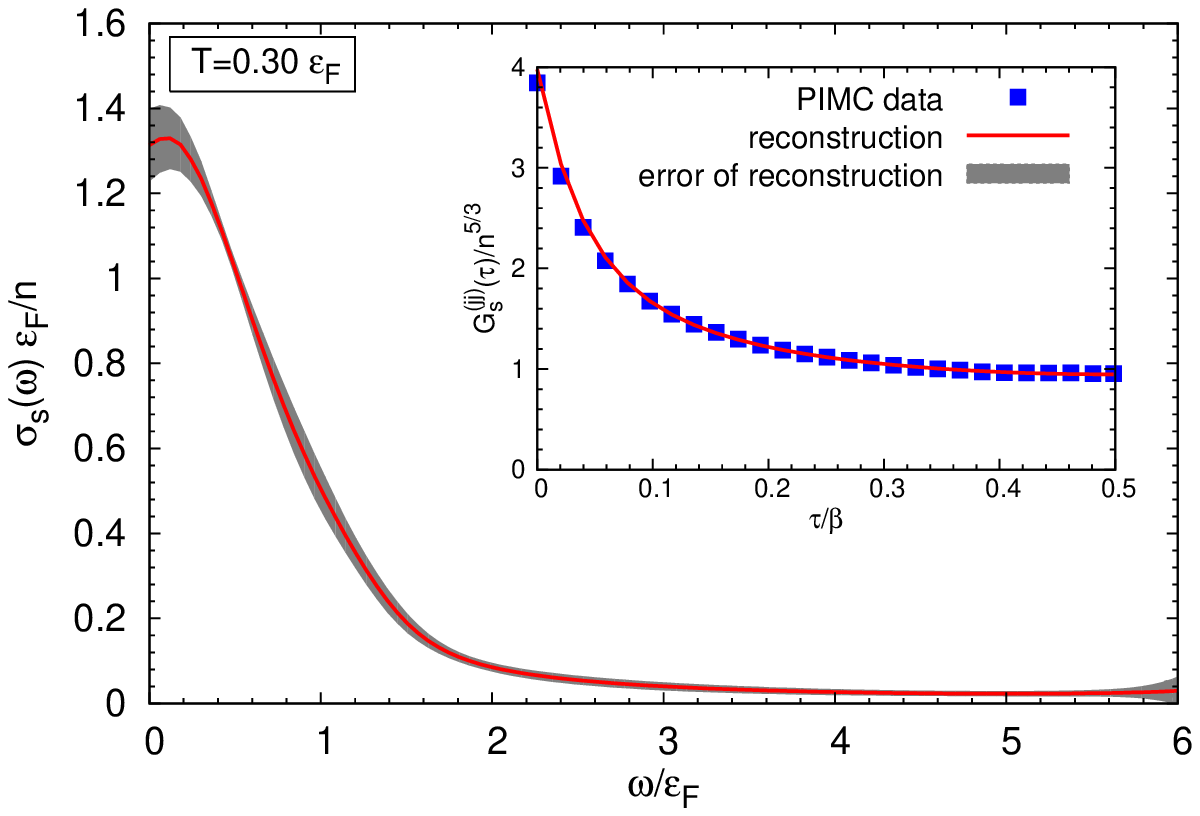}
\caption{ (Color online) The frequency dependent spin conductivity obtained from the Quantum Monte Carlo calculations with $10^3$ lattice for three different temperatures: upper panel $T=0.15\,\eF$ (at the critical temperature), middle panel $T=0.20\,\eF$ (pseudogap regime) and lower panel $T=0.30\,\eF$ (normal phase). The gray bands show uncertainty of the reconstruction. In the insets by (blue) points the corresponding PIMC correlators are presented. The line shows quality of reproduction of the PIMC data.
\label{fig:spin_cond} }
\end{figure}
The inversion is based on combination of two complementary methods:  Singular Value Decomposition (SVD) and self-consistent Maximum Entropy Method (MEM), described in details in paper~\cite{MagierskiWlazlowskiSupp}. The inversion procedure consists of two steps:
\begin{enumerate}
 \item the SVD solution $\tilde{\sigma}_{s}(\omega)$ is created. This step exploits only the information contained 
  in the correlator $G^{(jj)}_{s}(\tau)$.
 \item the self-consistent MEM produces final solution $\sigma_{s}(\omega)$, which is forced to satisfy external constraints 
  of the form:
 \begin{eqnarray}
  & P[\sigma_{s}^{(jj)}(\omega)]=\tilde{\sigma}_{s}^{(jj)}(\omega), &\\
  & \sigma_{s}^{(jj)}(\omega)=\dfrac{C}{3\pi\omega^{3/2}},\quad \textrm{for}\quad\omega>\omega_{\textrm{tail}},&
 \end{eqnarray} 
where $P$ stands for the projection operator onto the SVD subspace. To represent the MEM solution, we used a mesh of $120$ points uniformly distributed within interval $[0,\omcut]$.
\end{enumerate}
The self-consistent MEM needs to be supplemented by a class of \apriori models for the solution. The class has been defined as
\begin{eqnarray}
 M(\omega,\{\mu,\gamma,c,\alpha^{}_{1},\alpha^{}_{2}\}) = f(\omega,\{\alpha^{}_{1},\alpha^{}_{2}\})\,\dfrac{C}{3\pi\omega^{3/2}}\nonumber\\
 +[1\!-\!f(\omega,\{\alpha^{}_{1},\alpha^{}_{2}\})]\, \mathcal{L}(\omega,\{\mu,\gamma,c\}),
\end{eqnarray} 
where
\begin{equation}
  f(\omega,\{\alpha^{}_{1},\alpha^{}_{2}\})=e^{-\alpha^{}_{1}\alpha^{}_{2}}\dfrac{e^{\alpha^{}_{1}\omega}-1}{1+e^{\alpha^{}_{1}(\omega-\alpha^{}_{2})}},
\end{equation} 
and
\begin{equation}
 \mathcal{L}(\omega,\{\mu,\gamma,c\})=c\,\dfrac{1}{\pi} \, \frac{\gamma}{(\omega-\mu)^2 + \gamma^{2}}.
\end{equation} 
The parameters $\{\mu,\gamma,c,\alpha^{}_{1},\alpha^{}_{2}\}$  describe admissible degrees of freedom of
the model and are adjusted automatically in the self-consistent manner. To initialize the model for the first iteration 
we fit it to the SVD solution. 

The algorithm includes three ``bootstrap'' parameters:
i) $\omcut$ - the upper limit of the integration, 
ii) $\omega^{}_{\textrm{tail}}$ - the point where the universal tail behavior is imposed, 
iii) $\alpha$ - the parameter which governs the relative importance of the \apriori model to the data.
The bootstrap sample consists of about 200 launches of the algorithm with randomly generated (from some reasonable intervals) bootstrap parameters. Since each solution satisfy the integral equation~(\ref{eqn:InversionProblemSupp})  also the solution produced as an average over the collection satisfies it as well. This solution is regarded to be  the most favorable scenario, while the standard deviation over the bootstrap sample is associated with the uncertainty of the inversion procedure.

We have used the Lorentzian-like structure $\mathcal{L}(\omega)$ at low frequencies (Drude model) for the conductivity as it correctly reproduces $T$-matrix results for $\omega<\eF$ as reported in~\cite{EnssSupp}. Moreover we have checked the stability of the solution when one uses the Gaussian-like structure  instead of the Lorentzian. We have not found significant differences for the static spin conductivity $\sigma_s$ 
as we change the low frequency part of the model.

In~\fig{fig:spin_cond} we present a sample of results for three temperatures obtained for $10^3$ lattice. In the normal phase $T\gtrsim 0.25\,\eF$ we observe that $\sigma_s(\omega)$ possesses a Lorentzian-like structure at low frequencies with a maximum in the vicinity of zero frequency. Once we enter the pseudogap regime, the suppression of $\sigma_s(\omega)$ develops for the zero frequency, due to the onset of pairing correlations.

\end{document}